\documentclass[10pt]{article}
\usepackage{graphicx}
\usepackage{amsmath}
\usepackage{amssymb}
\usepackage{caption2}
\begin{document}
\begin{center}
{\large {\bf \sc{  Analysis of  the $X(5568)$ as scalar tetraquark state in the diquark-antidiquark model with QCD sum rules }}} \\[2mm]
Zhi-Gang Wang \footnote{E-mail: zgwang@aliyun.com.  }     \\
 Department of Physics, North China Electric Power University, Baoding 071003, P. R. China
\end{center}

\begin{abstract}
In this article, we take the $X(5568)$   as the  diquark-antidiquark type tetraquark state with the spin-parity $J^P=0^+$,  construct   the scalar-diquark-scalar-antidiquark type current, carry out the operator product expansion up to  the vacuum condensates of dimension-10, and study the mass and pole residue  in details with the QCD sum rules. We obtain the value  $M_X=\left(5.57\pm0.12 \right) \,\rm{GeV}$, which is consistent with the experimental data.  The present prediction favors assigning  the $X(5568)$ to be the scalar tetraquark  state.
\end{abstract}



\section{Introduction}

Recently,  the  D0 collaboration  observed   a narrow structure, $X(5568)$, in the decay chains  $X(5568) \to B_s^0 \pi^{\pm}$  ,$B_s^0 \to J/\psi
\phi$, $J/\psi\to \mu^+ \mu^-$  , $\phi \to K^+K^-$ with significance  of $5.1\sigma$ based on $10.4~\rm{fb^{-1}}$  of $p \bar{p}$  collision data at   $\sqrt{s}=1.96\,\rm{TeV}$  collected  at the Fermilab Tevatron collider \cite{X5568-exp}.   The mass and natural width of the new state are   $M_X =
5567.8 \pm 2.9 { }^{+0.9}_{-1.9}\,\rm{MeV} $   and $\Gamma_X = 21.9 \pm 6.4 {}^{+5.0}_{-2.5}\,\rm{MeV}$,  respectively.
The $B_s^0 \pi^{\pm}$ systems consist of two quarks and two antiquarks of four different flavors.
The D0 collaboration fitted the $B_s^0 \pi^{\pm}$ systems with the Breit-Wigner parameters in relative S-wave,  the favored  quantum numbers are $J^P =0^+$. However,
the quantum numbers  $J^P = 1^+$ cannot be excluded according to  decays  $X(5568) \to B_s^*\pi^+ \to B_s^0 \pi^+ \gamma$, where the low-energy photon is not
detected.

In this article, we assume the   $X(5568)$ to be the scalar diquark-antidiquark type  tetraquark state. There are five types  diquarks,   scalar diquarks,
 pseudoscalar diquarks, vector diquarks, axialvector diquarks and  tensor diquarks according to the structures  in Dirac spinor space. The
 favored configurations are the scalar diquarks  and axialvector diquarks  from the QCD sum rules \cite{WangDiquark,WangLDiquark}.  The   heavy scalar and axialvector  diquarks have almost  degenerate masses \cite{WangDiquark}, while   the masses of the
 light  axialvector  diquarks lie   $(150-200)\,\rm{MeV}$ above  the corresponding   light scalar diquarks \cite{WangLDiquark}.
  We take the  scalar light diquark and heavy diquark  as the basic constituents \cite{Maiani2004},  construct  the scalar-diquark-scalar-antidiquark type    current, which is expected to couple potentially to the lowest state, to study the mass and pole residue  of the $X(5568)$    with the QCD sum rules  \cite{SVZ79, PRT85}.
In the charm sector, the $D_s(2307)$ has been studied as the scalar-diquark-scalar-antidiquark type tetraquark state based on the QCD sum rules \cite{Ds2317}.

 The article is arranged as follows:  we derive the QCD sum rules for the mass and pole residue of  the
$X(5568)$ in Sect.2;  in Sect.3, we present the numerical results and discussions; and Sect.4 is reserved for our
conclusion.

\section{QCD sum rules for  the $ X(5568)$ as scalar  tetraquark state}

In the following, we write down  the two-point correlation function $\Pi(p)$ in the QCD sum rules,
\begin{eqnarray}
\Pi(p)&=&i\int d^4x e^{ip \cdot x} \langle0|T\left\{J(x)J^{\dagger}(0)\right\}|0\rangle \, ,
\end{eqnarray}
\begin{eqnarray}
   J(x)&=&\epsilon^{ijk}\epsilon^{imn}u^j(x)C\gamma_5 s^k(x) \bar{d}^m(x)\gamma_5 C \bar{b}^n(x) \, ,
\end{eqnarray}
where the $i$, $j$, $k$, $m$, $n$ are color indexes, the $C$ is the charge conjugation matrix.

We  insert  a complete set of intermediate hadronic states with the same quantum numbers as the current operator  $J(x)$ into the
correlation function  $\Pi(p)$ to obtain the hadronic representation \cite{SVZ79,PRT85}. After isolating the ground state
contribution of the scalar tetraquark state, we get the following result,
\begin{eqnarray}
\Pi(p)&=&\frac{\lambda_{ X}^2}{M_{X}^2-p^2} +\cdots \, \, ,
\end{eqnarray}
where the   pole residue  $\lambda_{X}$ is defined by $\langle 0|J(0)|X(p)\rangle = \lambda_{X}$ .

 In the following,  we carry out  the operator product expansion.  We contract the $u$, $d$, $s$ and $c$ quark fields in the correlation function $\Pi(p)$ with Wick theorem, and obtain the result:
\begin{eqnarray}
 \Pi(p)&=&i\epsilon^{ijk}\epsilon^{imn}\epsilon^{i^{\prime}j^{\prime}k^{\prime}}\epsilon^{i^{\prime}m^{\prime}n^{\prime}}\int d^4x e^{ip \cdot x}   \nonumber\\
&&{\rm Tr}\left[ \gamma_{5}S^{kk^{\prime}}(x)\gamma_{5} CU^{jj^{\prime}T}(x)C\right] {\rm Tr}\left[ \gamma_{5} B^{n^{\prime}n}(-x)\gamma_{5} C D^{m^{\prime}mT}(-x)C\right]   \, ,
\end{eqnarray}
where  the $U_{ij}(x)$, $D_{ij}(x)$, $S_{ij}(x)$ and $B_{ij}(x)$ are the full $u$, $d$, $s$ and $b$ quark propagators respectively (the $U_{ij}(x)$, $D_{ij}(x)$, $S_{ij}(x)$ can be written as $S_{ij}(x)$ for simplicity, where $q=u,\,d,\,s$.),
\begin{eqnarray}
S_{ij}(x)&=& \frac{i\delta_{ij}\!\not\!{x}}{ 2\pi^2x^4}-\frac{\delta_{ij}m_q}{4\pi^2x^2}-\frac{\delta_{ij}\langle
\bar{q}q\rangle}{12} +\frac{i\delta_{ij}\!\not\!{x}m_q
\langle\bar{q}q\rangle}{48}-\frac{\delta_{ij}x^2\langle \bar{q}g_s\sigma Gq\rangle}{192}+\frac{i\delta_{ij}x^2\!\not\!{x} m_q\langle \bar{q}g_s\sigma
 Gq\rangle }{1152}\nonumber\\
&& -\frac{ig_s G^{a}_{\alpha\beta}t^a_{ij}(\!\not\!{x}
\sigma^{\alpha\beta}+\sigma^{\alpha\beta} \!\not\!{x})}{32\pi^2x^2}  -\frac{1}{8}\langle\bar{q}_j\sigma^{\mu\nu}q_i \rangle \sigma_{\mu\nu}+\cdots \, ,
\end{eqnarray}
\begin{eqnarray}
B_{ij}(x)&=&\frac{i}{(2\pi)^4}\int d^4k e^{-ik \cdot x} \left\{
\frac{\delta_{ij}}{\!\not\!{k}-m_b}
-\frac{g_sG^n_{\alpha\beta}t^n_{ij}}{4}\frac{\sigma^{\alpha\beta}(\!\not\!{k}+m_b)+(\!\not\!{k}+m_b)
\sigma^{\alpha\beta}}{(k^2-m_b^2)^2}\right.\nonumber\\
&&\left. -\frac{g_s^2 (t^at^b)_{ij} G^a_{\alpha\beta}G^b_{\mu\nu}(f^{\alpha\beta\mu\nu}+f^{\alpha\mu\beta\nu}+f^{\alpha\mu\nu\beta}) }{4(k^2-m_b^2)^5}+\cdots\right\} \, ,\nonumber\\
f^{\alpha\beta\mu\nu}&=&(\!\not\!{k}+m_b)\gamma^\alpha(\!\not\!{k}+m_b)\gamma^\beta(\!\not\!{k}+m_b)\gamma^\mu(\!\not\!{k}+m_b)\gamma^\nu(\!\not\!{k}+m_b)\, ,
\end{eqnarray}
and  $t^n=\frac{\lambda^n}{2}$, the $\lambda^n$ is the Gell-Mann matrix  \cite{PRT85}, then compute  the integrals both in the coordinate space and momentum space to obtain the correlation function $\Pi(p)$ at the quark level, therefore the QCD spectral density through dispersion relation $\rho(s)=\frac{{\rm Im}\Pi(s)}{\pi}$. The explicit expression is neglected for simplicity. In Eq.(5), we retain the terms $\langle\bar{q}_j\sigma_{\mu\nu}q_i \rangle$  come from the Fierz re-arrangement of the $\langle q_i \bar{q}_j\rangle$ to  absorb the gluons  emitted from other quark lines   to extract the mixed condensates  $\langle\bar{q}g_s\sigma G q\rangle$, several new terms appear and play an important role in determining the Borel window.

 In this article, we carry out the operator product expansion up to the vacuum condensates  of dimension-10, and
assume  vacuum saturation for the  higher dimension vacuum condensates.
The condensates $\langle \frac{\alpha_s}{\pi}GG\rangle$, $\langle \bar{q}q\rangle\langle \frac{\alpha_s}{\pi}GG\rangle$, $\langle \bar{s}s\rangle\langle \frac{\alpha_s}{\pi}GG\rangle$, $\langle \bar{q}q\rangle^2\langle \frac{\alpha_s}{\pi}GG\rangle$,
$\langle \bar{s}s\rangle^2\langle \frac{\alpha_s}{\pi}GG\rangle$, $\langle \bar{q}q\rangle \langle \bar{s}s\rangle\langle \frac{\alpha_s}{\pi}GG\rangle$,
 $\langle \bar{q} g_s \sigma Gq\rangle^2$, $\langle \bar{q} g_s \sigma Gq\rangle\langle \bar{s} g_s \sigma Gs\rangle$ and $\langle \bar{s} g_s \sigma Gs\rangle^2$ are the vacuum expectations of the operators of the order $\mathcal{O}(\alpha_s)$. We take the truncations $n\leq 10$ and $k\leq 1$ in a consistent way,
the operators of the orders $\mathcal{O}( \alpha_s^{k})$ with $k> 1$ are  neglected.
  The condensates $\langle g_s^3 GGG\rangle$, $\langle \frac{\alpha_s GG}{\pi}\rangle^2$,
 $\langle \frac{\alpha_s GG}{\pi}\rangle\langle \bar{q} g_s \sigma Gq\rangle$ and $\langle \frac{\alpha_s GG}{\pi}\rangle\langle \bar{s} g_s \sigma Gs\rangle$ have no contributions.

 Once the spectral density at the quark level is obtained,  we can take the
quark-hadron duality below the continuum threshold  $s_0$ and perform Borel transform  with respect to
the variable $P^2=-p^2$ to obtain  the QCD sum rule:
\begin{eqnarray}
\lambda^2_{X}\, \exp\left(-\frac{M^2_{X}}{T^2}\right)= \int_{m_b^2}^{s_0} ds\, \rho(s) \, \exp\left(-\frac{s}{T^2}\right) \, ,
\end{eqnarray}
where
\begin{eqnarray}
\rho(s)&=&\frac{1}{6144\pi^6}\int_\Delta^1 dx\, x(1-x)^4 \left(s-\widetilde{m}_b^2\right)^3\left(3s-\widetilde{m}_b^2\right)\nonumber\\
&&-\frac{m_b\langle\bar{q}q\rangle}{64\pi^4}\int_\Delta^1 dx\, (1-x)^2\left(s-\widetilde{m}_b^2\right)^2\nonumber\\
&&-\frac{m_s\left[2\langle\bar{q}q\rangle-\langle\bar{s}s\rangle\right]}{64\pi^4}\int_\Delta^1 dx \, x(1-x)^2\left(s-\widetilde{m}_b^2\right)\left(2s-\widetilde{m}_b^2\right) \nonumber\\
&&-\frac{m_b^2}{9216\pi^4}\langle\frac{\alpha_sGG}{\pi}\rangle\int_\Delta^1 dx\, \frac{(1-x)^4}{x^2} \left(3s-2\widetilde{m}_b^2\right)\nonumber\\
&&+\frac{1}{1536\pi^4}\langle\frac{\alpha_sGG}{\pi}\rangle\int_\Delta^1 dx\, (1+2x)(1-x)^2 \left(s-\widetilde{m}_b^2\right)\left(2s-\widetilde{m}_b^2\right)\nonumber\\
&&-\frac{m_b\langle\bar{q}g_s\sigma Gq\rangle}{128\pi^4}\int_\Delta^1 dx\, \frac{(1-x)(1-3x)}{x}\left(s-\widetilde{m}_b^2\right) \nonumber\\
&&+\frac{m_s\left[3\langle\bar{q}g_s\sigma Gq\rangle-\langle\bar{s}g_s\sigma Gs\rangle\right]}{192\pi^4}\int_\Delta^1 dx\, x(1-x)\left(3s-2\widetilde{m}_b^2\right)\nonumber\\
&&+\frac{\langle\bar{q}q\rangle\langle\bar{s}s\rangle}{12\pi^2}\int_\Delta^1 dx\, x(1-x)\left(3s-2\widetilde{m}_b^2\right)\nonumber\\
&&+\frac{m_s m_b \langle\bar{q}q\rangle\left[2\langle\bar{q}q\rangle-\langle\bar{s}s\rangle\right]}{24\pi^2}\int_\Delta^1 dx\nonumber\\
&&-\frac{ m_b\langle\bar{q}q\rangle}{192\pi^2}\langle\frac{\alpha_sGG}{\pi}\rangle\int_\Delta^1 dx\, \frac{(1-x)^2}{x^2}-\frac{m_b\langle \bar{q}q\rangle}{144\pi^2}\langle\frac{\alpha_sGG}{\pi}\rangle\int_\Delta^1 dx \nonumber\\
&&+\frac{ m_b^3\langle\bar{q}q\rangle}{576\pi^2}\langle\frac{\alpha_sGG}{\pi}\rangle\int_\Delta^1 dx\, \frac{(1-x)^2}{x^3}\delta\left(s-\widetilde{m}_b^2\right)\nonumber\\
&&+\frac{m_s m_b^2\left[2\langle\bar{q}q\rangle- \langle\bar{s}s\rangle\right]}{1152\pi^2}\langle\frac{\alpha_sGG}{\pi}\rangle\int_\Delta^1 dx\, \frac{(1-x)^2}{x^2} \left(1+\frac{\widetilde{m}_b^2}{T^2} \right)\delta\left(s-\widetilde{m}_b^2\right)\nonumber\\
&&-\frac{m_s \left[2\langle\bar{q}q\rangle-\langle\bar{s}s\rangle\right]}{384\pi^2}\langle\frac{\alpha_sGG}{\pi}\rangle\int_\Delta^1 dx\,(1-x) \left[2+\widetilde{m}_b^2\left(s-\widetilde{m}_b^2\right)\right] \nonumber\\
&&-\frac{m_s\langle\bar{q}q\rangle}{576\pi^2}\langle\frac{\alpha_sGG}{\pi}\rangle\int_\Delta^1 dx\, x \left[2+\widetilde{m}_b^2\left(s-\widetilde{m}_b^2\right) \right]\nonumber\\
&&-\frac{ \langle\bar{q}q\rangle\langle\bar{s}g_s \sigma Gs\rangle+\langle\bar{s}s\rangle\langle\bar{q}g_s \sigma Gq\rangle}{48\pi^2}\int_\Delta^1 dx\, x \left[ 2+\widetilde{m}_b^2\left(s-\widetilde{m}_b^2\right)\right]\nonumber\\
&&-\frac{m_s m_b\left[ 12\langle\bar{q}q\rangle\langle\bar{q}g_s \sigma Gq\rangle-2\langle\bar{q}q\rangle\langle\bar{s}g_s \sigma Gs\rangle-3\langle\bar{s}s\rangle\langle\bar{q}g_s \sigma Gq\rangle\right]}{288\pi^2}   \delta\left(s-m_b^2\right)\nonumber\\
&&+\frac{m_s m_b\left[2\langle\bar{q}q\rangle-\langle\bar{s}s\rangle \right]\langle\bar{q}g_s\sigma Gq\rangle}{96\pi^2}\int_\Delta^1 dx\, \frac{1}{x}\delta\left(s-\widetilde{m}_b^2\right) \nonumber\\
&&-\frac{m_b\langle\bar{q}q\rangle^2\langle\bar{s}s\rangle}{9} \delta\left(s-m_b^2\right)+\frac{\langle\bar{q}g_s\sigma Gq\rangle\langle\bar{s}g_s\sigma Gs\rangle}{192\pi^2} \left(1+\frac{m_b^2}{T^2} \right)\delta\left(s-m_b^2\right)\nonumber\\
&&+\frac{m_s m_b\langle\bar{q}g_s\sigma Gq\rangle\left[\langle\bar{s}g_s\sigma Gs\rangle-3\langle\bar{q}g_s\sigma Gq\rangle \right]}{576 \pi^2 T^2} \left(1-\frac{m_b^2}{T^2} \right) \delta\left(s-m_b^2\right) \nonumber\\
&&+\frac{\langle\bar{q}q\rangle\langle\bar{s}s\rangle}{144}\langle\frac{\alpha_sGG}{\pi}\rangle\int_\Delta^1 dx\, \left(1+\frac{\widetilde{m}_b^2}{T^2}\right)\delta\left(s-\widetilde{m}_b^2\right) \nonumber\\
&&-\frac{ \langle\bar{q}q\rangle\langle\bar{s}s\rangle}{216T^4}\langle\frac{\alpha_sGG}{\pi}\rangle\int_\Delta^1 dx\, \frac{1-x}{x} \widetilde{m}_b^4 \delta\left(s-\widetilde{m}_b^2\right)\nonumber\\
&&+\frac{m_s m_b\langle\bar{q}q\rangle\left[2\langle\bar{q}q\rangle-\langle\bar{s}s\rangle\right]}{144T^2}\langle\frac{\alpha_sGG}{\pi}\rangle\int_\Delta^1 dx\, \frac{1}{x^2}\left(1-\frac{\widetilde{m}_b^2}{3T^2} \right)\delta\left(s-\widetilde{m}_b^2\right)\nonumber\\
&&+\frac{\langle\bar{q}q\rangle\langle\bar{s}s\rangle}{216}\langle\frac{\alpha_sGG}{\pi}\rangle \left(1+\frac{m_b^2}{T^2}\right)\delta\left(s-m_b^2\right) \nonumber\\
&&+\frac{m_s m_b^3\langle\bar{q}q\rangle\left[4\langle\bar{q}q\rangle-\langle\bar{s}s\rangle\right]}{864T^4}\langle\frac{\alpha_sGG}{\pi}\rangle \delta\left(s-m_b^2\right) \, ,
\end{eqnarray}
$\Delta=\frac{m_b^2}{s}$, $\widetilde{m}_b^2=\frac{m_b^2}{x}$, and $\int_\Delta^1 dx \to \int_0^1 dx$ when the $\delta \left(s-\widetilde{m}_b^2 \right)$ appears.

We differentiate   Eq.(7) with respect to  $\frac{1}{T^2}$, then eliminate the  pole residue $\lambda_{X}$, and obtain the QCD sum rule for
 the mass of the $X(5568)$,
 \begin{eqnarray}
 M^2_{X}= \frac{\int_{m_b^2}^{s_0} ds\frac{d}{d \left(-1/T^2\right)}\rho(s)\exp\left(-\frac{s}{T^2}\right)}{\int_{m_b^2}^{s_0} ds \rho(s)\exp\left(-\frac{s}{T^2}\right)}\, .
\end{eqnarray}

\section{Numerical results and discussions}
The input parameters are shown explicitly in Table 1.
The quark condensates, mixed quark condensates and $\overline{MS}$ masses  evolve with the   renormalization group equation, we take into account
the energy-scale dependence according to the following equations,
\begin{eqnarray}
\langle\bar{q}q \rangle(\mu)&=&\langle\bar{q}q \rangle(Q)\left[\frac{\alpha_{s}(Q)}{\alpha_{s}(\mu)}\right]^{\frac{4}{9}}\, , \nonumber\\
 \langle\bar{s}s \rangle(\mu)&=&\langle\bar{s}s \rangle(Q)\left[\frac{\alpha_{s}(Q)}{\alpha_{s}(\mu)}\right]^{\frac{4}{9}}\, , \nonumber\\
 \langle\bar{q}g_s \sigma Gq \rangle(\mu)&=&\langle\bar{q}g_s \sigma Gq \rangle(Q)\left[\frac{\alpha_{s}(Q)}{\alpha_{s}(\mu)}\right]^{\frac{2}{27}}\, , \nonumber\\ \langle\bar{s}g_s \sigma Gs \rangle(\mu)&=&\langle\bar{s}g_s \sigma Gs \rangle(Q)\left[\frac{\alpha_{s}(Q)}{\alpha_{s}(\mu)}\right]^{\frac{2}{27}}\, , \nonumber\\
m_b(\mu)&=&m_b(m_b)\left[\frac{\alpha_{s}(\mu)}{\alpha_{s}(m_b)}\right]^{\frac{12}{25}} \, ,\nonumber\\
m_s(\mu)&=&m_s({\rm 2GeV} )\left[\frac{\alpha_{s}(\mu)}{\alpha_{s}({\rm 2GeV})}\right]^{\frac{4}{9}} \, ,\nonumber\\
\alpha_s(\mu)&=&\frac{1}{b_0t}\left[1-\frac{b_1}{b_0^2}\frac{\log t}{t} +\frac{b_1^2(\log^2{t}-\log{t}-1)+b_0b_2}{b_0^4t^2}\right]\, ,
\end{eqnarray}
  where $t=\log \frac{\mu^2}{\Lambda^2}$, $b_0=\frac{33-2n_f}{12\pi}$, $b_1=\frac{153-19n_f}{24\pi^2}$, $b_2=\frac{2857-\frac{5033}{9}n_f+\frac{325}{27}n_f^2}{128\pi^3}$,  $\Lambda=213\,\rm{MeV}$, $296\,\rm{MeV}$  and  $339\,\rm{MeV}$ for the flavors  $n_f=5$, $4$ and $3$, respectively  \cite{PDG}.
 Furthermore, we set the $u$ and $d$ quark masses to be zero.

\begin{table}
\begin{center}
\begin{tabular}{|c|c|c|c|}\hline\hline
    Parameters                                          & Values\\   \hline
   $\langle\bar{q}q \rangle({\rm 1GeV})$                & $-(0.24\pm 0.01\, \rm{GeV})^3$ \,\, \cite{SVZ79,PRT85,ColangeloReview}         \\  \hline
   $\langle\bar{s}s \rangle({\rm 1GeV})$                & $(0.8\pm0.1)\langle\bar{q}q \rangle({\rm 1GeV})$ \,\, \cite{SVZ79,PRT85,ColangeloReview}     \\ \hline
$\langle\bar{q}g_s\sigma G q \rangle({\rm 1GeV})$       & $m_0^2\langle \bar{q}q \rangle({\rm 1GeV})$   \,\,  \cite{SVZ79,PRT85,ColangeloReview}       \\  \hline
$\langle\bar{s}g_s\sigma G s \rangle({\rm 1GeV})$       & $m_0^2\langle \bar{s}s \rangle({\rm 1GeV})$  \,\,  \cite{SVZ79,PRT85,ColangeloReview}        \\  \hline
$m_0^2({\rm 1GeV})$                                     & $(0.8 \pm 0.1)\,\rm{GeV}^2$      \,\,  \cite{SVZ79,PRT85,ColangeloReview}    \\   \hline
 $\langle \frac{\alpha_s GG}{\pi}\rangle$               & $(0.33\,\rm{GeV})^4$          \,\,  \cite{SVZ79,PRT85,ColangeloReview} \\   \hline
   $m_{b}(m_b)$                                         & $(4.18\pm0.03)\,\rm{GeV}$ \,\, \cite{PDG}      \\    \hline
   $m_{s}({\rm 2GeV})$                                  & $(0.095\pm0.005)\,\rm{GeV}$  \,\, \cite{PDG}      \\   \hline \hline
\end{tabular}
\end{center}
\caption{ The  input parameters in the QCD sum rules, the values in the bracket denote the energy scales $\mu=1\,\rm{GeV}$, $2\,\rm{GeV}$ and $m_b$, respectively. }
\end{table}

In Refs.\cite{Wang-tetraquark-cc, Wang-tetraquark-bb}, we study the acceptable energy scales of the QCD spectral densities  for the hidden  charm (bottom) tetraquark states  in the QCD sum rules in details for the first time,  and suggest a  formula $\mu=\sqrt{M^2_{X/Y/Z}-(2{\mathbb{M}}_Q)^2}$ to determine  the energy scales, where the $X$, $Y$, $Z$ are the four-quark systems, and the ${\mathbb{M}}_Q$ are the effective heavy quark masses. The energy scale formula has been successfully extended to study the charmed baryon states and hidden-charm  pentaquark states \cite{WangBP}. Recently, we re-checked the numerical calculations and found that there exists  a small error involving the mixed condensates  \cite{Wang-tetraquark-bb}.  The Borel windows are modified slightly and the numerical results are also improved slightly after the small error is corrected, the conclusions survive, the optimal value of the effective mass is ${\mathbb{M}}_b=5.17\,\rm{GeV}$  instead of $5.13\,\rm{GeV}$ for  the hidden-bottom tetraquark states. In this article, we use the energy scale formula $\mu=\sqrt{M^2_{X}-\mathbb{M}_b^2}$ to determine the optimal energy scale and obtain the  value $\mu=2.1\,\rm{GeV}$.

In this article, the QCD spectral density  $\rho(s)\propto s^n$ with $n\leq 4$,  the   integral $\int_0^{\infty} \rho(s) \exp \left(-\frac{s}{T^2}\right) ds$ converges
slowly, it is difficult to obtain the pole contribution larger than $50\%$.   In calculations, we use the QCD spectral density $\rho(s)\theta(s-s_0)$
to approximate the continuum contribution with the value $\sqrt{s_0}=6.1\pm0.1\,\rm{GeV}$, where we take it for granted that the energy gap between the ground state and the first radial excited state is about $0.4-0.6\,\rm{GeV}$, just like the conventional mesons and the hidden-charm tetraquark states, the  $Z(4430)$ is assigned to be  the first radial excitation of the $Z_c(3900)$ \cite{Maiani-Nielsen,Wang3900-4430}. Now we  search for the optimal  Borel parameter $T^2$ according to the two criteria (pole dominance  and convergence of the operator product expansion) of the QCD sum rules.
 The optimal Borel parameter $T^2=(4.5-4.9)\,\rm{GeV}^2$, the pole contribution is about $(15-29)\%$. At the operator product expansion side, the contributions of the vacuum condensates $D_i$ of dimension $i$ have the relations,  $D_3, \,D_6,\,|D_8|, \,D_9\gg D_0,\,D_4,\,|D_5|,\,D_7,\,D_{10}$, $D_0+D_3\approx 40\%$,
  $D_4+D_5\approx -1\%$, $D_6\approx 66\%$, $D_7+D_8+D_9\approx-10\%$, $D_{10}\approx 5\%$. The dominant contributions come from the terms $D_0+D_3+D_6$, the operator product expansion is convergent, but the pole contribution is smaller than $30\%$, we cut down the continuum contamination by the threshold parameter $s_0$. The radial excited states or high resonances have to  be included in, if one wants to obtain QCD sum rules with the pole contributions larger than $50\%$ \cite{WangNPA2007}.

We take into account  all uncertainties  of the input   parameters,
and obtain the values of the mass and pole residue of
 the $X(5568)$  as the scalar diquark-antidiquark type tetraquark state,
\begin{eqnarray}
M_X&=&\left(5.57\pm0.12 \right) \,\rm{GeV} \, , \nonumber \\
\lambda_X&=&\left(6.7\pm 1.6\right)\times 10^{-3} \,\rm{GeV}^5 \, .
\end{eqnarray}
  In Fig.1, we plot the predicted mass with variation of the Borel parameter. From the figure, we can see that the  platform is rather flat, and we expect to make reasonable prediction.   The predicted mass $M_X=\left(5.57\pm0.12 \right) \,\rm{GeV}$ is consistent with the experimental data $M_X =
5567.8 \pm 2.9 { }^{+0.9}_{-1.9}\,\rm{MeV} $ from the D0 collaboration \cite{X5568-exp}.

\begin{figure}
 \centering
 \includegraphics[totalheight=6cm,width=8cm]{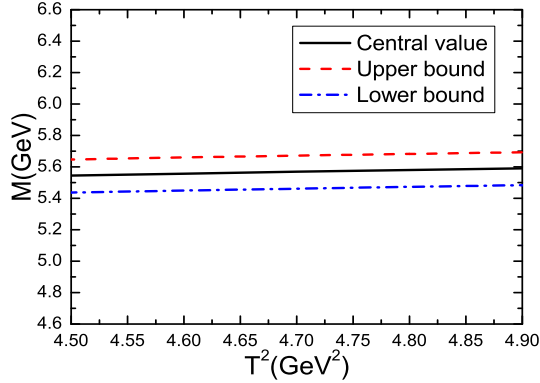}
         \caption{ The mass  $M_X$  with variation of the Borel parameter $T^2$.  }
\end{figure}

In Ref.\cite{Nielsen5568}, Zanetti,  Nielsen and  Khemchandani choose  the same interpolating  current as the present work, while in Ref.\cite{Azizi5568},   Agaev, Azizi and Sundu choose the axialvector-diquark-axialvector-antidiquark type scalar  current.
If we apply the identity $\epsilon^{ijk}\epsilon^{imn}=\delta^{jm}\delta^{kn}-\delta^{jn}\delta^{km}$ in the color space to the current $J(x)$, we can obtain the current used in Ref.\cite{ChenZhu5568} in studying  the $X(5568)$ as a scalar tetraquark state with the QCD sum rules. In Ref.\cite{ChenZhu5568}, Chen et al also study the $X(5568)$ as the axialvector tetraquark state and reproduce the experimental value of the mass $M_{X(5568)}$.  The quantum numbers  $J^P = 1^+$ cannot be excluded according to  decays  $X(5568) \to B_s^*\pi^+ \to B_s^0 \pi^+ \gamma$, where the low-energy photon is not
detected.
The present article and Refs.\cite{Nielsen5568,Azizi5568,ChenZhu5568} appeared in the net http://arxiv.org/ on the same day, and the calculations were done independently. In Refs.\cite{Azizi5568,ChenZhu5568}, the parameters in Table 1 are taken directly to evaluate the QCD spectral densities,   while in the present work we evolve the values to a special energy scale $\mu=2.1\,\rm{GeV}$ determined by the energy scale formula. The input parameters chosen in Ref.\cite{Nielsen5568} are  different  from the values  used in the present work and in Refs.\cite{Azizi5568,ChenZhu5568}.
The experimental value of the mass $M_{X(5568)}$ can be well reproduced  in the present work and in Refs.\cite{Nielsen5568,Azizi5568,ChenZhu5568}.

In the following, we  perform Fierz re-arrangement to the current $J$ both in the color and Dirac-spinor  spaces to obtain the  result,
\begin{eqnarray}
J&=&\frac{1}{4}\left\{\,-\bar{b} s\,\bar{d} u+\bar{b}i\gamma_5 s\,\bar{d}i\gamma_5 u-\bar{b} \gamma^\mu s\,\bar{d}\gamma_\mu u
-\bar{b} \gamma^\mu\gamma_5 s\,\bar{d}\gamma_\mu\gamma_5 u+\frac{1}{2}\bar{b}\sigma_{\mu\nu} s\,\bar{d}\sigma^{\mu\nu} u\right. \nonumber\\
&&\left.+\bar{b} u\,\bar{d} s-\bar{b}i\gamma_5 u\,\bar{d}i\gamma_5 s+\bar{b} \gamma^\mu u\,\bar{d}\gamma_\mu c+
\bar{b} \gamma^\mu\gamma_5 u\,\bar{d}\gamma_\mu\gamma_5 s-\frac{1}{2}\bar{b}\sigma_{\mu\nu} u\,\bar{d}\sigma^{\mu\nu} s  \,\right\} \, ,
\end{eqnarray}
where $\bar{b}\Gamma s\,\bar{d}\Gamma u=\bar{b}^{j}\Gamma s^{j}\,\bar{d}^{k}\Gamma u^{k}$ and $\bar{b}\Gamma u\,\bar{d}\Gamma s=\bar{b}^{j}\Gamma u^{j}\,\bar{d}^{k}\Gamma s^{k}$, $\Gamma=1$, $i\gamma_5$, $\gamma_\mu$, $\gamma_\mu\gamma_5$, $\sigma_{\mu\nu}$, the $j$ and $k$ are color indexes.
The components $\bar{b}i\gamma_5 s\,\bar{d}i\gamma_5 u$ and $\bar{b} \gamma^\mu\gamma_5 s\,\bar{d}\gamma_\mu\gamma_5 u$ couple potentially   to the meson pair   $B_s^0\pi^{+}$, while the components $\bar{b}i\gamma_5 u\,\bar{d}i\gamma_5 s$ and $\bar{b} \gamma^\mu\gamma_5 u\,\bar{d}\gamma_\mu\gamma_5 s$ couple potentially   to the meson pair   $B^+\bar{K}^0$. The strong decays
\begin{eqnarray}
X(5568) &\to& B_s^0\pi^{+} \, ,
\end{eqnarray}
are Okubo-Zweig-Iizuka  super-allowed, while the decays
\begin{eqnarray}
X(5568) &\to& B^+\bar{K}^{0} \, ,
\end{eqnarray}
  are kinematically forbidden, which is consistent with the observation of the D0 collaboration \cite{X5568-exp}. The present work favors assigning the $X(5568)$ to be the scalar   diquark-antidiquark type tetraquark state.

In Ref.\cite{Swanson5568},  Burns  and Swanson  argue that it is unusual to assign the $X(5568)$ to be the threshold effect, the $B_s^*\pi-B_s\pi$ cusp with $J^P=1^-$, the $B_s\pi-B\bar{K}$ cusp with $J^P=0^+$, the Gamow-Gurney-Condon type resonance induced by the $B_s\pi-B\bar{K}$ interaction, the compact tetraquark state based on  phenomenological analysis. In Ref.\cite{Guo5568},    Guo, Meissner and Zou  provide additional arguments using the chiral symmetry and heavy quark symmetry. According to the constituent quark model, the constituent diquark model, the  chiral symmetry and heavy quark symmetry, the lowest mass of the $su\bar{b}\bar{d}$ tetraquark state with $J^P=0^+$ is much larger than the $M_{X(5568)}$ \cite{Swanson5568,Guo5568,WangZhu5568}, for more references on this subject, one can consult Refs.\cite{Swanson5568,Guo5568}. In the present work and in Refs.\cite{Nielsen5568,Azizi5568,ChenZhu5568}, the $X(5568)$ is assigned to be the compact tetraquark state, the experimental value of the mass $M_{X(5568)}$ can be well reproduced based on the QCD sum rules.  In Ref.\cite{Decay5568},  the partial decay width of the strong decay  $ X(5568) \to B_s^0 \pi^+$ is studied with the three-point QCD sum rules. If we saturate the width of the $X(5568)$ with  the strong decay  $ X(5568) \to B_s^0 \pi^+$, the experimental value $\Gamma_X = 21.9 \pm 6.4 {}^{+5.0}_{-2.5}\,\rm{MeV}$ can be reproduced approximately. More theoretical works and more experimental data are still needed to approve existence or non-existence of the $X(5568)$.

\section{Conclusion}
In this article, we take the $X(5568)$  to be the scalar  diquark-antidiquark type tetraquark state,  construct   the scalar-diquark-scalar-antidiquark type current,
 carry out the operator product expansion up to  the vacuum condensates of dimension-10, and study the mass and pole residue of the $X(5568)$ in details with the QCD sum rules.   In calculations,  we use the  formula $\mu=\sqrt{M^2_{X}-{\mathbb{M}}_b^2}$  to determine  the energy scale of the QCD spectral density.  The present prediction favors assigning  the $X(5568)$  to be the scalar tetraquark  state. The  pole residue can be taken as   basic input parameter to study relevant processes of the $X(5568)$ with the three-point QCD sum rules.

\section*{Acknowledgements}
This  work is supported by National Natural Science Foundation,
Grant Number 11375063, and Natural Science Foundation of Hebei province, Grant Number A2014502017.

\end{document}